\documentstyle[epsf,psfig]{mn}

\title{Stokes imaging, Doppler mapping and Roche tomography of the AM Her
system V834 Cen }

\author[Stephen \,B. Potter et al.] {Stephen \,B. Potter,$^{1}$, E.
Romero--Colmenero$^{1}$, C. A. Watson$^{2}$, D. A. H. Buckley$^{1}$ \&
A. Phillips$^{3}$\\ 
$^{1}$South African Astronomical Observatory, PO Box 9,
Observatory 7935, Cape Town, South Africa\\ 
$^{2}$Department of Physics and Astronomy, University of Sheffield, Sheffield
S3 7RH\\
$^{3}$Pergamentum Solutions, 409 Bonner Lane, Bozeman, MT 59715}

\newcommand{\eg}{{\it e.g.\ }}

\date{}

\begin{document}

\maketitle

\begin{abstract}

We report on new simultaneous phase resolved spectroscopic and
polarimetric observations of the polar (AM Herculis star) V834~Cen
during a high state of accretion. Strong emission lines and high
levels of variable circular and linear polarization are observed over
the orbital period.

The polarization data is modelled using the Stokes imaging technique of Potter
et al. The spectroscopic emission lines are investigated using the Doppler
tomography technique of Marsh \& Horne and the Roche tomography technique of
Dhillon and Watson.

Up to now all three techniques have been used separately to investigate the
geometry and accretion dynamics in Cataclysmic Variables. For the first time,
we apply all three techniques to simultaneous data for a single system. This
allows us to compare and test each of the techniques against each other and
hence derive a better understanding of the geometry, dynamics and system
parameters of V834 Cen.

All three techniques are consistent with an interpretation in which a
ballistic stream extends to a minimum of $\sim$ 40 degrees in azimuth
around the white dwarf before becoming threaded by the magnetic field
lines. Interestingly, the observed ballistic Doppler velocites do not
show a reduced $v_{y}$ component, as found in Doppler imaging of other
AM Her systems. Furthermore, the secondary star in V834 Cen shows more
He II (4686 \AA \ ) emission on its leading inner face, as opposed to
the trailing face like in other AM Her systems. We propose that the
accretion shock preferentially illuminates the leading face of the
secondary star. In addition, the ballistic stream does not
obscure the leading face of the secondary from the accretion shock,
and, in fact, our Doppler maps show that the ballistic stream is a
strong He II (4686 \AA \ ) emission source in itself and thus adds to
the illumination of the leading face of the secondary.

\end{abstract}

 \begin{keywords}
     accretion, accretion discs -- methods: analytical -- techniques:
     polarimetric -- binaries: close -- novae, cataclysmic variables --
     X--rays: stars. 
 \end{keywords}

\section{Introduction}

The standard picture of a cataclysmic variable (CV) is a binary system
in which there is a Roche-lobe filling red dwarf (the secondary star)
and an accreting white dwarf (the primary), see \eg Warner 1995 for a
review of Cataclysmic Variables. In AM Her systems, also known as
polars, the white dwarf primary has a sufficiently strong magnetic
field to lock the system into synchronous rotation and to prevent an
accretion disc from forming.  Instead, the material overflowing the
Roche lobe of the secondary initially continues on a ballistic
trajectory until, at some distance from the white dwarf, the magnetic
pressure overwhelms the ram pressure of the ballistic stream. From
this point onwards the accretion flow is confined to follow the
magnetic field lines of the white dwarf, see \eg Cropper 1990 or
Warner 1995 for a full review of magnetic CVs.

The now supersonic accreting material suddenly becomes subsonic at a
shock region, which forms at some height above the white dwarf
surface. The shock-heated material reaches temperatures of $\sim$
10-50 keV and is ionised. The hot plasma cools by two different
mechanisms as it settles onto the white dwarf, resulting in a density
and temperature stratification in the post-shock region. The two
cooling mechanisms are X-ray cooling, in the form of bremsstrahlung
radiation and, with sufficient magnetic fields, cyclotron cooling in
the form of optical/IR cyclotron radiation.

V834~Cen was discovered as an X--ray source (E1405-451 = H1405-45) with the
HEAO 1 and Einstein satellites by Jensen et al. (1982), and Mason et al. (1993)
identified it with a stellar object. It was later given the classification of
an AM Her system when Tapia (1982) and Bailey et al. (1983) reported strong
linear and circular phase dependent polarization. Extensive polarimetric and
spectroscopic observations by Cropper et al. (1986) and Rosen et al. (1987)
followed.

In 1987, Wickramasinghe, Tuohy \& Visvanathan detected the presence of
strongly phase-dependent absorption features in the spectrum of V834
Cen. They interpreted the features as Zeeman components of H$\alpha$,
originating from a cloud of cool gas surrounding the accretion
shock. Puchnarewicz et al. (1990) and Schwope \& Beuermann (1990) also
reported the presence of pure photospheric Zeeman absorption lines
during a low state. Ferrario et al. (1992) also reported observations
of V834~Cen when it was in a low state. They detected Zeeman
absorption lines from the photosphere of the white dwarf and cyclotron
emission features from the accretion shock, indicating that there was
still some accretion from the companion star. These authors were
therefore able to unambiguously determine the magnetic field strength
and structure of V834~Cen, which, in turn, makes it an ideal object
for Stokes imaging (Potter, Hakala \& Cropper 1998).

Mauche (2002) reported on EUVE (Extreme Ultraviolet Explorer)
observations of V834~Cen during two epochs of different accretion
states. He gave a qualitative explanation for the EUV light curves by
invoking a simple model of accretion from a ballistic stream along
field lines of a tilted magnetic field centered on the white
dwarf. During the higher accretion state, accretion would occur over a
broader range of azimuths than during the lower accretion state.

In this paper, a new version of Stokes imaging, which was described in
Potter, Hakala \& Cropper (1998) in its original form, is applied to
polarimetric observations of V834~Cen. Here we also describe our new,
more robust, methodology for Stokes imaging, which allows objective
mapping of the cyclotron emission regions in magnetic cataclysmic
variables in terms of their location, shape and size.  In addition, we
apply Doppler mapping, of Marsh \& Horne (1988), to spectroscopic data
taken simultaneously and four months later than the polarimetry. The
Doppler solutions are then modelled using a single particle trajectory
code and compared to the results derived from Stokes imaging. Finally
we apply Roche tomography, of Dhillon \& Watson (2001) to the same
spectroscopic data in order to gain knowledge of the irradiation
pattern of the inner irradiated surface of the companion star and
compare the results with those obtained from the other two techiques.

\section{Observations}

\begin{table*}
\begin{center}
\caption{Log of V834 Cen observations. 1.9m and 1.0m are at the
SAAO. `Cass spect' is the Cassegrain spectrograph. `UCTPol' is the
University of Cape Town photopolarimeter. `Orbits' is the number of
orbits each observation covered.  {\label{tab:observations}}}
\vspace{0.2cm}
\centerline{ 
\begin{tabular}{|l|c|l|l|c|c|c|} \hline
Date & Telescope & Instrument & Spectral range/filter & Resolution & Orbits & No. Spectra \\
\\ 
8/9 Apr 2000   & 1.9m & Cass spect       &3900 - 5500 \AA & 1.4 \AA (G6)  & 1.5   & 38 \\
8/9 Apr 2000   & 1.0m & UCTPol     & Clear           & -             & 1.5   & - \\
9/10 Apr 2000   & 1.0m & UCTPol     & Clear           & -             & 3.5   & - \\
10/11 Apr 2000   & 1.9m & Cass spect     &3900 - 5500 \AA & 1.4 \AA (G6) & 4.1   & 108 \\
10/11 Apr 2000   & 1.0m & UCTPol     & Clear           & -             & 3.5   & - \\
1/2 Aug 2000   & 1.9m & Cass spect     &4200 - 5000 \AA & 0.5 \AA (G4) & 1.6& 22 \\
5/6 Aug 2000   & 1.9m & Cass spect     &4200 - 5000 \AA & 0.5 \AA (G4) & 2.0   & 37 \\
\\
\hline\hline 
\end{tabular}
}
\end{center}
\end{table*}

\subsection{Polarimetry data}

V834~Cen was observed on three nights during April 2000 (see
table~\ref{tab:observations}) using the South African Astronomical Observatory
(SAAO) 1.0-m telescope with the UCT polarimeter (UCTPol; Cropper 1985). The
UCTPol was operated in Stokes mode, i.e. simultaneous linear and circular
polarimetry, and photometry were obtained. White light observations were
undertaken, defined by an RCA31034A GaAs photomultiplier response 3500 -- 9000
\AA. Polarization standard stars (Hsu \& Breger 1982) were observed in order to
determine the position angle offsets. In addition non-polarized standard stars
and calibration polaroids were also observed in order to set the efficiency
factors. The data were reduced as described in Cropper (1997). A total of 8.5
orbits were observed.

\subsection{Spectroscopy data}
Optical spectroscopic observations were obtained on 2000 April and
August (see table ~\ref{tab:observations}) at the SAAO using the
1.9--m telescope, equipped with the Cassegrain spectrograph and
utilizing the SITe1 CCD ($1752 \times 266 \times 15\mu m$
pixels). Comparison arc (Cu-Ar) spectra were taken at regular
intervals. Flat fields and spectroscopic standards were taken on each
night in order to flux calibrate the data after wavelength
calibration. The data have been extinction corrected. A total of 9.2
orbits were observed.

\section{Analysis}

\subsection{Stokes imaging}

Stokes imaging reconstructs images of the cyclotron emission region on
the white dwarf by optimizing model fits to the intensity and
polarization light curves. The original Stokes imaging used models
based on the cyclotron emission calculations of Wickramasinghe \&
Meggitt (1985), which assumed homogeneous plasmas of constant electron
temperature and optical depth parameter. The model calculated the
viewing angle to any emitting point on the white dwarf and
interpolated on the Wickramasinghe \& Meggitt cyclotron models in
order to give the Stokes parameters ($I, Q, U$ and $V$) of the
radiation emitted by the point for all cyclotron harmonics. Emissions
from extended and/or multiple regions were calculated by summing the
emission from several such emission points. The light curves were then
constructed by viewing the ensemble of points as a function of phase.

The optimisation of the model to the data then proceeded by adjusting
the number and distribution of emission points across the surface of
the white dwarf. A genetic algorithm (GA, see for instance Charbonneau
1995) was used in order to optimise the fit. The GA first generated a
set of random solutions and calculated their ``fitness''. The fitness
of an individual solution was simply the sum of its $\chi^{2}$ fit to
the data plus a regularisation term, which was a measure of the
smoothness of the image solution. The GA then proceeded to minimise
the fitness of the solutions by ``breeding'' new solutions from the
best solutions produced so far. Eventually the improvement in the
fitness of the GA would level out and a best solution would be found.

\subsubsection{New methodology}

Since the conception and realization of Stokes imaging, several
aspects of the technique have been changed and/or improved in
order to make it more robust. We outline these below before we
describe its application to our new polarimetric observations.

Recently, new cyclotron emission calculations have been performed
which make use of more realistic shock structures (see Potter et
al. 2002). These new calculations better describe the observed
cyclotron spectral characteristics of mCVs, in particular the
continuum and the cyclotron humps. We have therefore replaced the
Wickramasinghe \& Meggitt (1985) calculations with these new
calculations at the core of Stokes imaging procedure. This is
particularly important when modelling multi-band polarimetric
observations.

We also addressed several other aspects of Stokes imaging. As can be
seen from figure 2 of Potter et al. (2001) for example, the fit to the
data appears remarkably good in some places and not so good in
others. The aim of Stokes imaging is to fit the general morphological
variations of the observations, but not to fit the small scale details
that either cannot be distinguished from noise or that go beyond the
assumptions of the cyclotron model. This was mostly controlled by the
choice of the regularisation term described above (i.e. the $\Lambda$
parameter in equation 2 of Potter et al. 1998). By adjusting
$\Lambda$, the smoothness of the image and, in turn, the goodness of
the fit to the data, could be changed. However, the choice of
$\Lambda$ was somewhat arbitrary and, as a result, the smoothness of
the emission region was predetermined; thus the technique was not
objective in this respect. In particular, it was unclear whether the
variation in brightness across the predicted emission region (see e.g.
Potter et al 2001) was truly a representation of the real emission
region or a result of over-fitting noisy parts of the
dataset. However, without the regularisation term, the number of
possible solutions was almost infinitely large.

We removed this ambiguity by eliminating the regularisation term from
our calculation of the fitness of the solutions, leaving the
$\chi^{2}$ term only. Instead, the technique is now constrained to
produce and breed solutions that consist of a {\it single} emission
region of any shape, size and location. We also allow for brightness
variations across the emission region. Optimisation is now performed
on a parameterized equation that describes the shape, size and
location of the emission region. Therefore, the size of the emission
region is no longer predetermined. Although the technique is
constrained to find solutions that contain a single emission region,
it is possible to allow more regions, for example for systems that
exhibit two pole accretion.

After running the technique several times, each starting with a different set
of randomly generated solutions, we found that in each case the final solution
was slightly different. However, there are only subtle differences in the fits,
as one solution may have fit one part of the light curves slightly better than
another solution.  All the image solutions predict a region at the same
location with roughly the same shape and size, and there are only subtle
differences in the brightness distribution from one image to the next. Hence,
it is clear that each execution of the technique is finding a local minimum in
the multi-dimensional parameter space, but they all are located quite close
within the global minimum.

Therefore, instead of choosing one image as the final solution, we now
construct an image from the top 10 percent (based on their $\chi^{2}$) of our
final solutions by simply taking their average. This average image now
represents the most probable solution for the shape, size and location of the
cyclotron emission region. Similarly, the model light curves are constructed by
taking the average of the top 10 percent solutions.

With the original version of the code, the final image solutions often
showed low level stray pixels in addition to the main emission region,
which probably arose as a result of Stokes imaging trying to model
noisy parts of the observations. Therefore, in order to minimise the
amount of noise and/or flickering variations in the polarimetric light
curves, a box car function was applied in order to smooth the data
prior to the application of Stokes Imaging. If possible, the data are
also binned and averaged over several orbits. This has effectively
reduced to zero the number of low-level stray pixels in our final
images.

\subsubsection{Application to new polarimetric observations of V834~Cen}

\begin{figure}
\epsfxsize=8.5cm
\epsffile{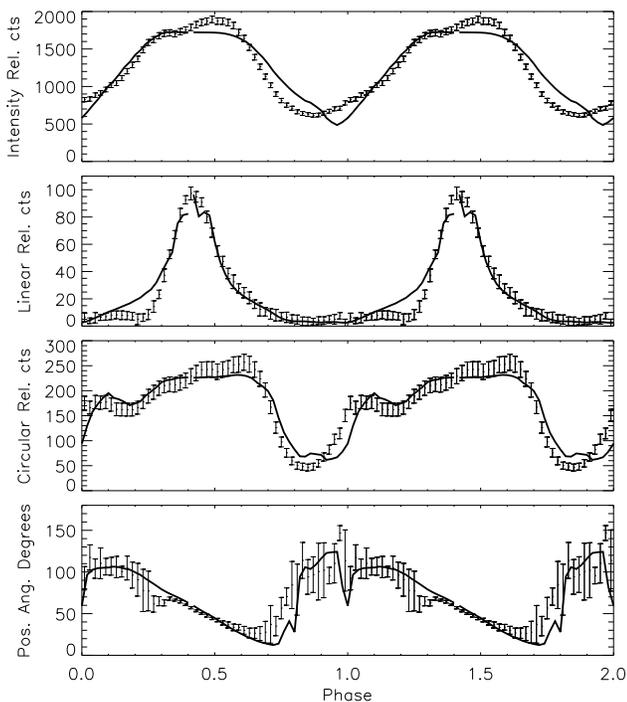} 
\caption{The polarimetric data (white light) folded on the
spectroscopic ephemeris of Schwope et al. (1993) together with the
model fit. }
\label{polfit}
\end{figure}

Fig.~\ref{polfit} shows the new polarimetric observations, spanning a
total of 8.5 orbits, folded and binned on the orbital ephemeris of
Schwope et al. (1993). As mentioned above, a box car function has also
been applied to the folded data in order to minimise the amount of
noise and/or flickering variations in the polarimetric light curves.

\begin{figure*}
\epsfxsize=17.5cm
\psfig{file=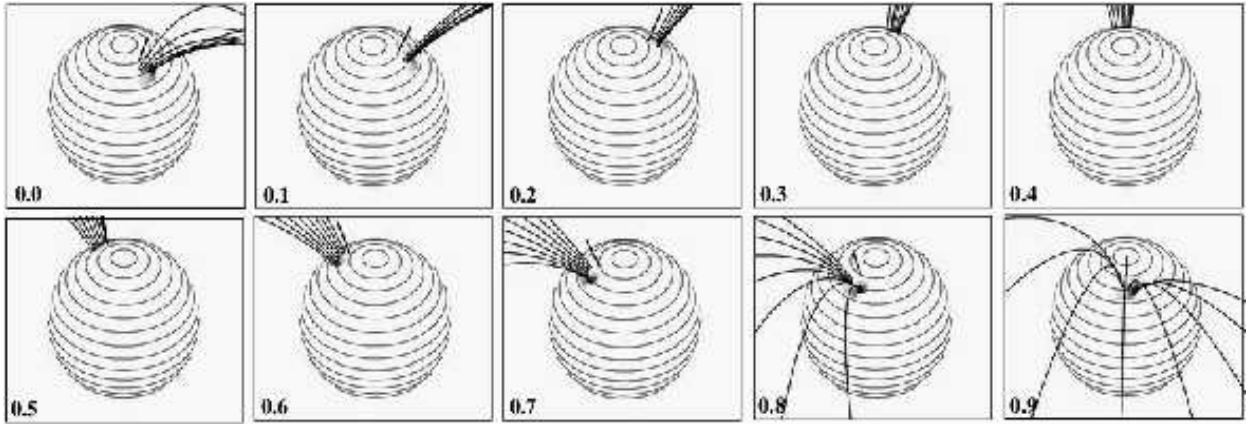,width=16.5cm,angle=-90}
\caption{View from the Earth of the white dwarf in V834~Cen for a complete
orbital rotation. The predicted cyclotron emission region is shown as the grey
scale region. The magnetic pole in indicated by a short solid line at latitude
10 degrees from the spin pole. Latitudes are indicated every 10
degrees. Magnetic field lines (see section 4.1) are shown as solid lines.  }
\label{globes}
\end{figure*}

Each execution of Stokes imaging assumes a fixed set of values for the
system parameters. From reviewing the literature (see introduction) we
have found a range of estimates for the inclination and dipole offset
angles of V834~Cen, thus we execute Stokes imaging several times in
order to investigate the parameter space. In Fig.~\ref{polfit} we show
the model fit assuming the system to have an inclination of 50 degrees
and dipole offset angles (latitude and longitude) of 20 and 36 degrees
respectively. Following Ferrario et al. (1992) the dipole is offset by
-0.1$R_{WD}$, giving a polar magnetic field strength of 31 MG The
choice of these parameters is discussed later (see discussion
section).

As can be seen from Fig.~\ref{polfit}, Stokes imaging has reproduced
the general morphological polarimetric variations very
well. Fig.~\ref{globes} shows the predicted shape, size and location
of the cyclotron emission region, together with some magnetic field
lines which will be discussed later.  As can be seen from this figure,
Stokes imaging predictes a region located approximately 10-15 degrees
from the upper magnetic pole. It is somewhat extended in longitude,
with most of its emission arising from the leading end of the
region. An enlarged view of the emission region can be seen in the
insert in Fig.~8. The emission region remains in view throughout the
whole orbit, thus the polarization light curves can be explained as a
combination of cyclotron beaming and self absorption by the accretion
shock. During phases 0.7 - 0.9, the emission region is most face on to
the viewer. As a result there is a dip in the circular polarization
and intensity at these phases due to cyclotron self absorption and
absorption by the accretion stream/column. During phases 0.2 - 0.5,
the emission region is seen on the far horizon of the white dwarf,
resulting in large amounts of linear polarization as we view the
emission region most perpendicular to the magnetic field lines that
feed it. The general variation of the position angle is also
accurately described by the model as simply due to the changing
viewing angle to the emission region as the white dwarf rotates.

\subsection{Doppler tomography}

\begin{figure}
\epsfxsize=8.5cm
\epsffile{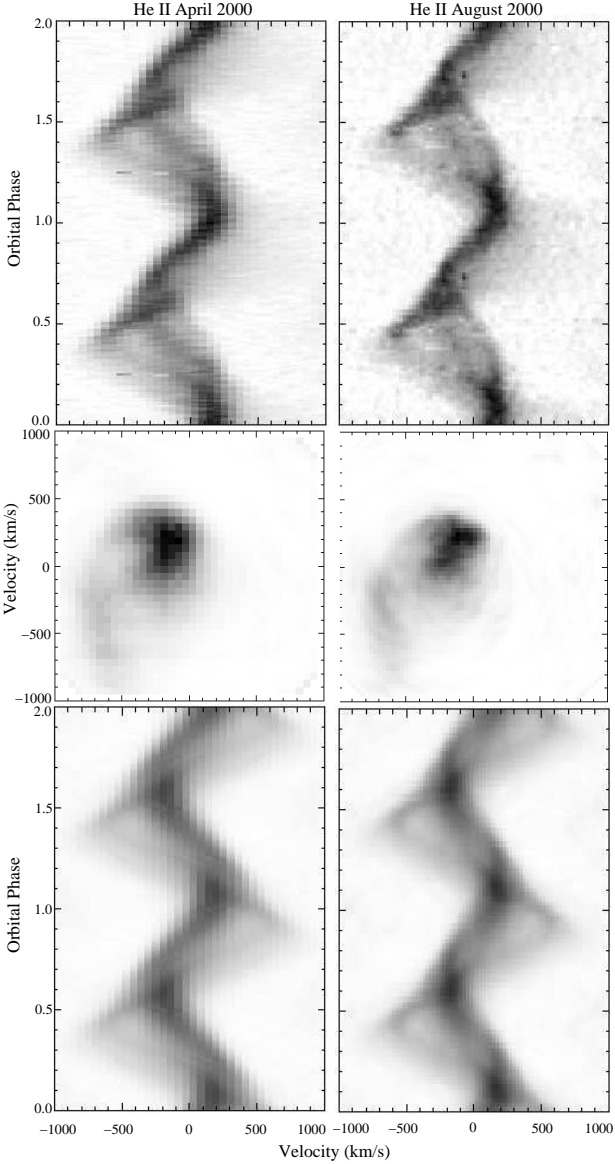} 
\caption{From top to bottom: HeII (4686 \AA) trailed spectra,
corresponding Doppler map and reconstructed trailed spectra. From left
to right: April 2000 (1.4 \AA resolution) and August 2000 (0.5 \AA
\ resolution) observations. See section 3.2 for details}
\label{trailed}
\end{figure}

\begin{figure}
\epsfxsize=8.5cm
\epsffile{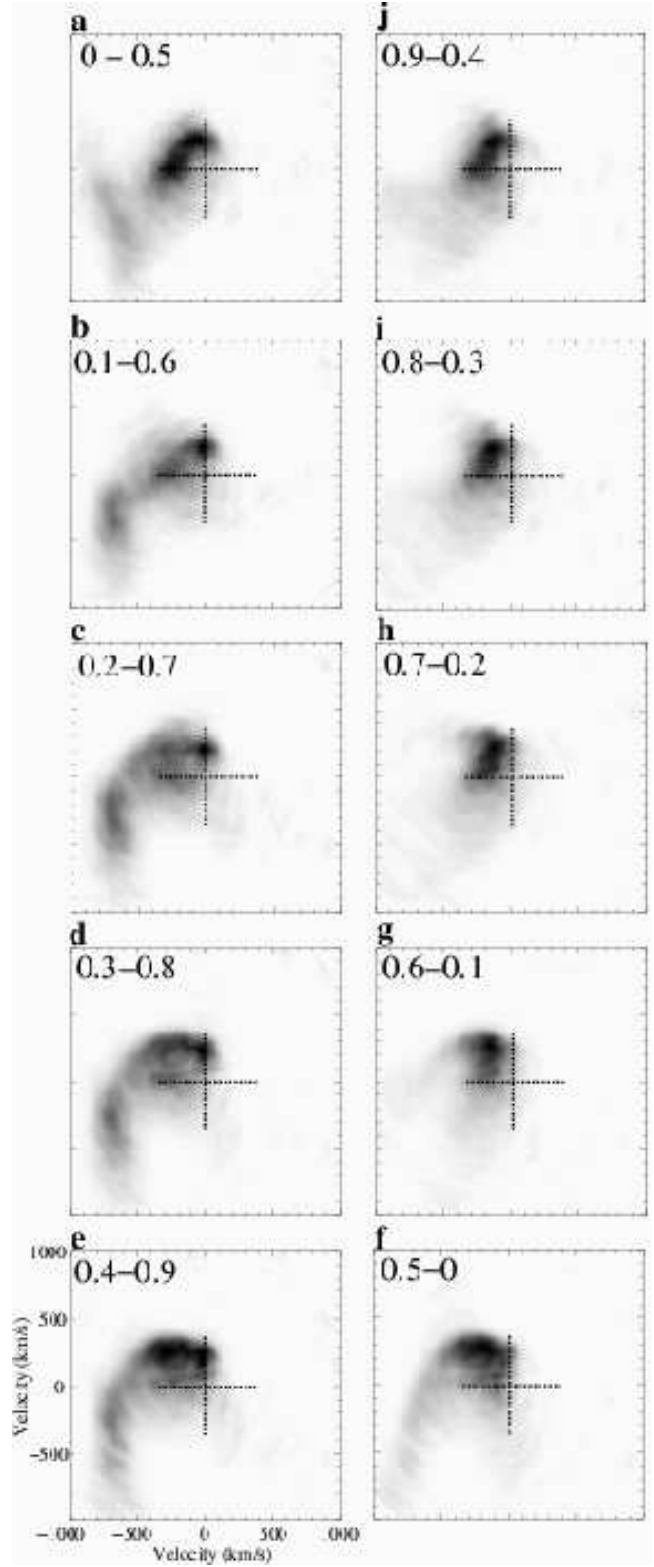} 
\caption{Doppler tomograms (Spruit code 1998) constructed using spectra taken
from consecutive half orbits. Plots are arranged in an anticlockwise direction
starting from the top left. See section 3.2. for details}
\label{anim}
\end{figure}

Fig.~\ref{trailed} (top-left) shows trailed HeII 4686 \AA \ emission
lines taken simultaneously with the polarimetry presented above, while
the top-right plot shows the trailed HeII spectra taken four months
later at higher wavelength resolution. Included in Fig.~\ref{trailed},
the Doppler tomograms (Spruit 1998) obtained from these data (middle
plots), and the synthetic trailed spectra (bottom), are shown. An
analysis of the two sets of observations gives the same
results. Therefore, in what follows, we present our analysis and
interpretation of the higher quality August 2000 observations.

Upon close inspection of the trailed spectra, one can see that it
consists of at least three components. This is borne out in the
tomograms, which show emission at the expected location of the
secondary star and/or the inner Lagrangian (L1) point, the ballistic
stream and perhaps parts of the magnetically confined flow. However,
the brightness scale of the tomogram is dominated by emission from the
secondary star and from around the L1 point. Consequently, the
accretion stream is somewhat difficult to discern.

In Fig.~\ref{anim} we show Doppler tomograms constructed using spectra
taken from consecutive half orbits. Here we are taking advantage of
the violation of the first axiom of Doppler tomography (Marsh 2001),
namely that Doppler tomography assumes that all points in the binary
system are equally visible at all times. This means that it is
possible to construct a tomogram using spectra covering half of an
orbit only, though in our case we obviously violate this axiom, as
most emission is optically thick. Consequently the tomograms of
Fig.~\ref{anim} are all different. By using half orbit spectra only,
we selectively eliminate various components of the binary system from
the tomogram, thus allowing otherwise less obvious features to become
more enhanced. For example, the irradiated face of the secondary star
is most visible at phase 0.5, hence the tomograms that were
constructed using spectra from around phase 0.5 quite clearly show
emission from the expected location of the secondary star
(Figs.~\ref{anim}a-e), and those that were not constructed around
phase 0.5 do not show the secondary (Figs.~\ref{anim}f-j).

\begin{figure*}
\epsfxsize=16.cm
\psfig{file=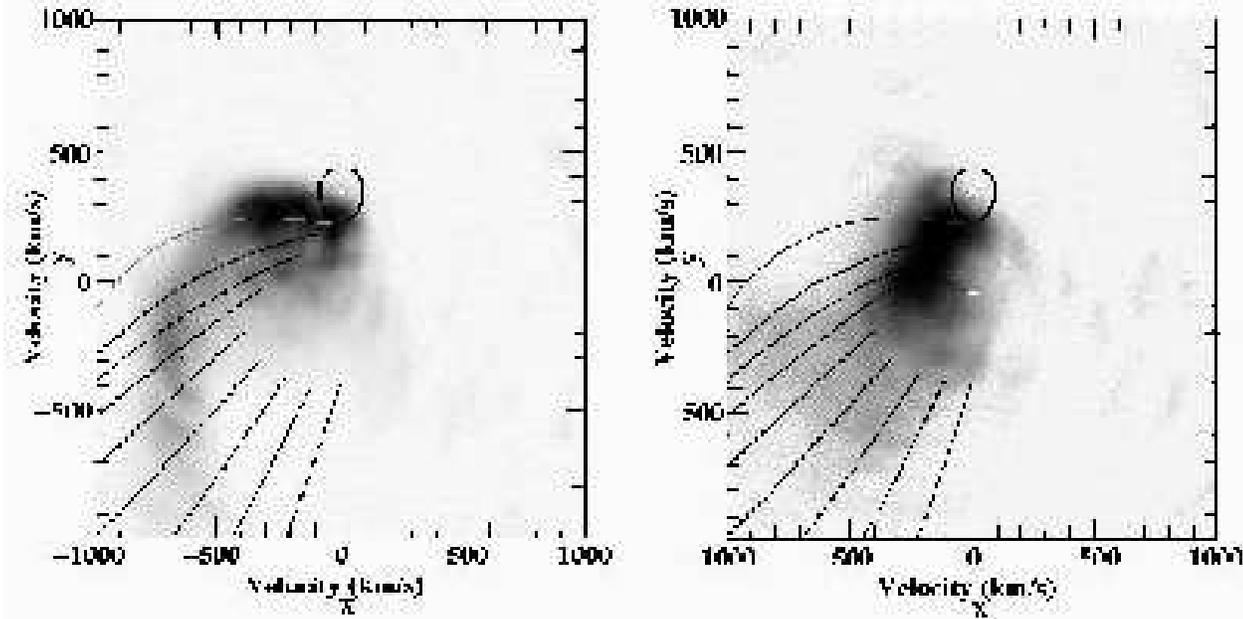,width=16.5cm,angle=-90}
\caption{Left and right plots show the Doppler tomograms reproduced
from Figs.~\ref{anim}e and \ref{anim}j respectively. Overplotted are
single particle ballistic trajectories and magnetic trajectories for
every 10 azimuthal degrees along the ballistic stream.}
\label{ballistic}
\end{figure*}

It is now possible to discern four and possibly five emission
components. From Fig.~\ref{anim}e we can see emission
at the expected location of the secondary star v$_{\rm x} \sim 0$,
v$_{\rm y} \sim 200$ km~s$^{-1}$, the ballistic stream (the ridge of
brightness starting at v$_{\rm y} \sim 200$ km~s$^{-1}$ and extending
from v$_{\rm x} \sim 0$ km~s$^{-1} $ to v$_{\rm x} \sim -600$
km~s$^{-1} $) and emission at a location that may be consistent with
part of the magnetic stream (region of emission centered on v$_{\rm y}
\sim$ -300, v$_{\rm x} \sim -700$ km~s$^{-1}$). In addition,
Fig.~\ref{anim}j shows a fourth component, consistent with the part of
the magnetic stream that has just left the orbital plane (the diagonal
region extending from v$_{\rm x} \sim 0$, v$_{\rm y} \sim 200$
km~s$^{-1}$ to v$_{\rm x} \sim -300$, v$_{\rm y} \sim 0$
km~s$^{-1}$. Furthermore, there is a fainter but broad emission
feature that occupies most of the lower left-hand quadrant of the
Doppler tomogram, which can be seen more clearly in the right plot of
Fig.~\ref{ballistic}. Therefore, in what follows, we will use the
tomograms of Fig.~\ref{anim}e and \ref{anim}j, which have been
reproduced in Fig.~\ref{ballistic}, in order to model and interpret
the dynamical components of the binary system.

\subsubsection{Single particle trajectory}

Our next step is to model the ballistic and magnetic streams of the
tomograms in order to gain a quantitative understanding of their
trajectories and to locate the footprints of the magnetic field lines
on the surface of the white dwarf. We can then compare the location of
the magnetic field lines with the location of the cyclotron emission
region predicted from Stokes imaging.

On the tomograms of Fig.~\ref{ballistic} (reproduced from
Figs.~\ref{anim}e and \ref{anim}j) we have overplotted a model
trajectory computed using a single particle under gravitational and
rotational influences, shown as the upper curved lines. No additional
drag terms (\eg due to the magnetic field) are included. The particle
is allowed to follow a ballistic path from the L1 point and is
terminated 75 degrees in azimuth around the white dwarf. At 10 degree
intervals in azimuth (from 5 to 75 degrees) around the white dwarf,
dipole trajectories are calculated (the straighter diagonal lines
below the ballistic trajectory) from the ballistic stream to the
surface of the white dwarf. The first dipole trajectory starts close
to the location of the secondary star, with the consecutive
trajectories starting at locations progressively closer to the white
dwarf. The velocity of the particle in the direction of the field is
conserved when it attaches onto a field line. At the threading region,
the re-direction of the particle is quite large in Doppler velocity
space, hence the beginning of the magnetic streams appear disjointed
from the ballistic stream.

The Doppler velocities are dependent on the inclination of the binary
system, the magnetic dipole angles and the mass ratio of the
binary. We used the same values as those for Stokes imaging, namely an
inclination of 50 degrees and dipole offset angles (latitude and
longitude) of 20 and 36 degrees respectively, consistent with those
calculated by Ferrario et al. (1992) and Schwope et al. (1993). The
magnetic dipole was again offset by -0.1 white dwarf radii along the
dipole axis, consistent with the findings of Ferrario et al. (1992).
A mass ratio $M_{1}/M_{2} = 6.5$ and white dwarf mass of 0.85M\sun \
were used, consistent (within the errors) with Schwope et al. 1993,
($M_{wd}\sun = 0.66^{+0.19}_{-0.16}$) but somewhat higher than
estimates from X-ray observations (Ramsay 2000, $M_{wd}\sun =
0.68$). The center of mass of the secondary star and the primary star
are indicated by a `+' and a `*' respectively. Also indicated is the
Roche lobe of the secondary star.

With this geometry, the single particle ballistic stream traces the
lower part of the emission fairly well (left panel
Fig.~\ref{ballistic}). Interestingly, the observed ballistic stream
does not appear to be dislocated to lower $v_{y}$ Doppler velocities
as observed in other systems. This is discussed later in the
discussion section.  The model ballistic trajectory is calculated up
to more than 40 degrees further than the end of the ballistic emission
seen in the tomogram.

The diagonal region extending from v$_{\rm x} \sim 0$, v$_{\rm y} \sim
200$ km~s$^{-1}$ to v$_{\rm x} \sim -300$, v$_{\rm y} \sim 0$
km~s$^{-1}$ (Fig.~\ref{ballistic} right plot) corresponds to the low
Doppler velocities of the threading region, from which most the
calculated magnetic trajectories point to.

The dipole trajectories also overlap with the emission seen centered
on v$_{\rm x} \sim$ -700 km~s$^{-1} $ and v$_{\rm y} \sim$ -300
km~s$^{-1} $ (best seen in the left plot of Fig.~\ref{ballistic}).
According to the calculated dipole trajectories, this area corresponds
to the highest point above the orbital plane where the dipole
trajectory ``turns over'', but see section 4.1.

The broad fainter emission, seen in the lower left quadrant of the
Doppler tomograms (best seen in the right plot of
Fig.~\ref{ballistic}), appears to correspond to the full range of
calculated dipole trajectories (5-75 degrees), suggesting that
threading of the magnetic field lines occurs from a ballistic stream
that extends by more than 40 degrees further than the Doppler
ballistic emission suggests.

However, it should be noted that the inclination of V834~Cen is
thought not to be very high, thus the out of plane velocities of the
dipole trajectory would lead to a smearing effect of the corresponding
emission in the Doppler tomograms. Thus a more detailed modelling than
presented here could lead to over interpretation.

\subsection{Roche tomography}

Roche tomography (Rutten \& Dhillon 1994; Watson \& Dhillon 2001;
Dhillon \& Watson 2001) is a technique used for imaging the
Roche-lobe-filling secondary stars in CVs. The secondary star is
modelled as a series of quadrilateral surface elements. Each element
is assigned its own local intrinsic specific intensity profile, which
are scaled to take into account the projected area, limb darkening and
obscuration. Every element is then Doppler shifted to the radial
velocity of the surface element at that particular phase. All the
elements are then summed to give the rotationally broadened profile at
any particular orbital phase. By iteratively varying the strengths of
the profile contributed from each element, the `inverse' of the above
procedure can be performed and thus Roche tomograms present images of
the distribution of line flux on the secondary star.

Due to the variable and unknown contribution to the spectrum of the
accretion region in V834~Cen, the data should ideally be slit-loss
corrected and continuum subtracted. The observations of V834~Cen have
not been slit-loss corrected, but we have folded the observations,
which cover several orbits, into twenty phase bins on the orbital
period, thus smearing out any effects due to random slit losses. In
addition we note that the transparency and seeing conditions were
excellent during all the observing runs and the slit-losses are
thought to be minimal. The data have been continuum subtracted.

We next fit Gaussians to the HeII 4686 \AA \ emission line profile in
each of the phase bins. During fitting, the solution from the previous
bin was used as a starting point for each consecutive bin. From these
fits we then extracted the parameters of the Gaussian that describes
the narrow component of the HeII emission line, which is thought to
arise from the irradiated face of the secondary. We then selected data
from only those phases where the narrow component is clearly
distinguishable from the other components. This is necessary in order
to minimise any artifacts that will arise in the Roche map as a result
of contamination of the input data from other components in the binary
system (e.g. from the accretion stream).

\begin{figure*}
\psfig{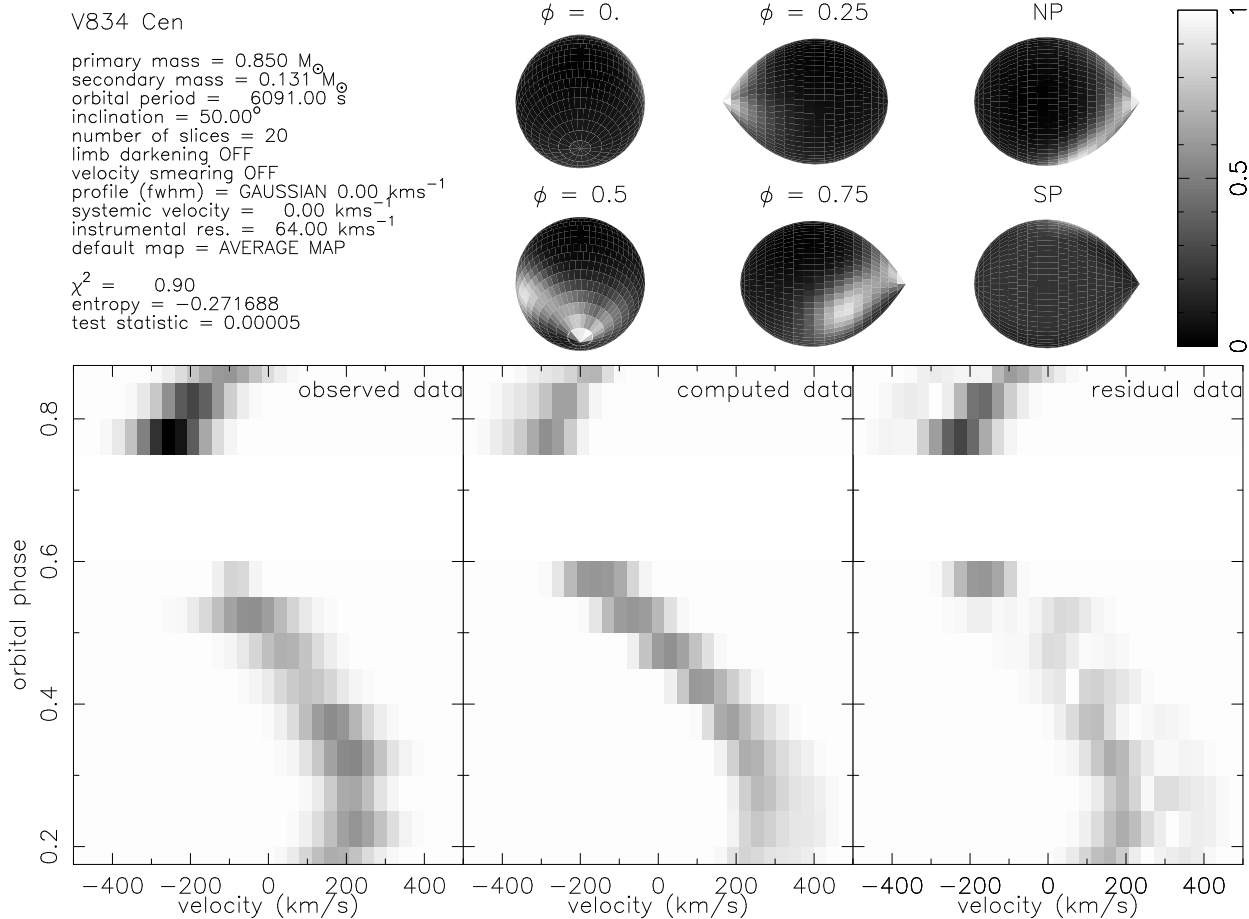}
\caption{Roche tomogram showing the distribution of HeII (4686 \AA)
emission on the Roche lobe of the secondary star in V834~Cen.}
\label{roche}
\end{figure*}

The Roche tomogram of V834~Cen obtained from the HeII 4686 \AA \
emission line is presented in Fig.~\ref{roche}. Also shown are the
input data and the computed data. There are two prominent asymmetries
evident in the tomograms. Firstly, a comparison of the Roche surface
presented at phases 0.25 and 0.75 reveals the leading face of the
secondary to be brighter than the trailing face. Secondly, the view of
the Roche surface from its north pole reveals that the inner surface
(the face of the secondary most pointing towards the white dwarf) is
brighter than the outer surface. These asymmetries can be explained as
a result of irradiation. This is discussed further in section 4.

\section{Discussion}

\begin{figure}
\epsfxsize=8.5cm
\epsffile{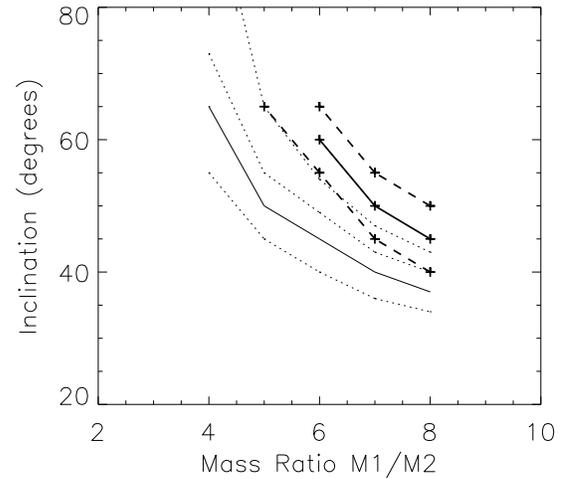} 
\caption{Lower thin solid and three dotted curves: The implied
inclination as a function of mass ratio for four values of the
effective radial velocity $k^{'}_{2} = 210,225,240$ and $260 kms^{-1}$
from bottom to top respectively reproduced from figure 5. of Schwope
et al. (1993). Upper thick solid curve shows the implied inclination
and mass ratios for a single particle ballistic trajectory to coincide
with the observed Doppler ballistic trajectory in
Fig.~\ref{ballistic}. The lower and upper dashed curves correspond to
single particle ballistic trajectories that are just lower and just
higher respectively than the observed Doppler ballistic trajectory in
Fig.~\ref{ballistic}.}
\label{syspar}
\end{figure}

\subsection{The system parameters}

As explained above, the calculated Doppler trajectory of the ballistic
stream depends on the system parameters, namely the inclination and
the mass ratio of the binary. We show in Fig.~\ref{syspar} the results
of calculating ballistic trajectories for a range of system parameters
and comparing them to the observed ballistic emission seen in the
tomograms. The thicker, upper solid curve of Fig.~\ref{syspar}
represents the range of system parameters that produce ballistic
streams which trace the center of the observed ballistic Doppler
emission fairly well.  The lower and upper dashed curves define system
parameters that give ballistic trajectories that lie just below and
above the observed ballistic emission respectively.

The lower four curves of Fig.~\ref{syspar} are reproduced from figure
5 of Schwope et al. (1993) and show their predictions for the system
parameters of V834~Cen based on modelling the $k^{'}_{2}$ velocities
of the secondary. More details of their model can be found in Beuermann
\& Thomas (1990). The dotted curves represent different estimates of
$k^{'}_{2}$.

From Fig.~\ref{syspar}, it is therefore immediately obvious that the
ballistic trajectories generally predict higher inclinations and/or
mass ratios than $k^{'}_{2}$ modelling. This is wholy unexpected, as
previous investigations of other systems such as HU Aqr (Schwope et
al. 1997, Heerlein et al. 1999) and QQ Vul (Schwope et al. 2000) have
shown the opposite to be true (see also the discussion by Schwope
2001). The main reason for previous findings was thought to be that
the, so-called, ballistic trajectory is not purely ballistic, but
suffers from a magnetic influence. This may either introduce a
magnetic drag to the ballistic particles and therefore slow them down
(see Sohl \& Wynn 1999), or the magnetic field is perhaps diverting
the stream to lower $v_{y}$ Doppler velocities. Our V834~Cen stream
appears to be more ballistic and/or less affected by magnetic
influences.

The overlap between the two models is minimal, with the center of the
overlap region being located at an inclination of 50 degrees with a
mass ratio of 6.5, giving a white dwarf mass of 0.85M\sun. The fit
obtained with Roche tomography, using the adopted parameters, (i.e. an
inclination of 50 degrees and a mass ratio of 6.5) predicts slightly
higher $k^{'}_{2}$ velocities than observed. Therefore Roche
tomography prefers lower inclinations and/or mass ratios which are,
not surprisingly, more inline with the $k^{'}_{2}$ velocity work of
Schwope et al. (1993). Stokes imaging is independent of mass ratio,
but it does predict a best inclination of 50 degrees (although the
larger range of 45-60 degrees cannot be excluded).

A mass ratio of 6.5 is relatively large amongst the magnetic
Cataclysmic Variables, and perhaps a possible reason for a true
ballistic stream. The high mass ratio may be enough to give the
ballistic stream sufficient ram pressure to overcome the magnetic
field of the white dwarf for at least some distance. We also point out
that our observations were taken when V834~Cen was in a high state,
implying a higher accretion rate and thus the ballistic stream would
have even more ram pressure than usual. Once the ballistic stream has
become sufficiently close to the white dwarf, its magnetic field
begins to dominate the flow of the stream, and the remaining ballistic
stream may even disappear from the Doppler tomogram as the velocities
become smeared. Another possibility may be that the magnetic field
geometry is such that its influence over the ballistic stream is not
very strong at the beginning of the ballistic trajectory.

In section 3.2 we briefly mentioned that the bright emission region
centered on v$_{\rm y} \sim$ -300, v$_{\rm x} \sim -700$ km~s$^{-1}$
may correspond to the highest point above the orbital plane where the
dipole trajectory ``turns over''. It may, however, be of ballistic
origin rather than magnetic origin. From the range of system
parameters defined by the models in Fig.~\ref{syspar}, it is not
possible to 'make' the simple, single particle ballistic stream go
through this region. However, we expect that adding an arbitrary
amount of magnetic drag could deflect the ballistic trajectory through
this region. Support for a ballistic stream that continues further
than the Doppler map suggests can further be obtained from EUVE
(Extreme UltraViolet Explorer) observations (Mauche 2002). That
analysis placed the EUV emission at a location on the white dwarf,
approximately 40 degrees in azimuth, extending to over 70 degrees when
V834~Cen is observed in a high state.

\subsection{The accretion region}

As mentioned in section 3.1, Fig.~\ref{globes} shows the white dwarf
of V834~Cen for a complete orbital rotation, as viewed from Earth. In
this spatial coordinate frame we have also plotted the magnetic field
lines that were derived from the modelling of the Doppler tomograms
(section 3.2). Upon close examination it is clear that footprints of
the magnetic field lines on the surface of the white dwarf coincide
fairly well with the location of the cyclotron emission region
predicted from Stokes imaging. In addition, the general longitudinal
extent of the cyclotron emission runs along the footprints of the
magnetic field lines. Furthermore, the brightest part of the cyclotron
emission region coincides with the magnetic field lines connecting
with the end of the ballistic stream, thus adding further support that
magnetic accretion does occur towards the end of the ballistic
stream.

As already mentioned above, the EUV emission emanates from a similar
location on the surface of the white dwarf (see Mauche 2002).

\subsection{Irradiation of the secondary}

The Roche tomogram displayed in Fig.~\ref{roche} shows that the HeII
emission distribution to be stronger on the leading face of the
secondary star. A close inspection of the tomograms in
Figs.~\ref{anim}b-d also indicate a possible asymmetry towards the
leading face of the secondary.  This is contrary to other magnetic
Cataclysmic Variables for which HeII Roche tomography is available.

It is generally accepted that the HeII emission from the Roche surface of the
secondary is as a result of irradiation by the accretion shock. Watson et
al. (2003) found for three magnetic CVs (AM Her, QQ Vul and HU Aqr), that the
irradiation pattern (from HeII emission and Na I absorption roche tomograms)
was mostly on the trailing faces of the secondary stars. They reasoned that the
accretion stream/column was somehow shielding the leading face of the secondary
from the shock on the surface of the white dwarf. 

In the case of V834~Cen we must, therefore, reason that any shielding
is somehow minimal. Instead, the leading face of the secondary is more
irradiated because of the simple reason that, from its position within
the orbital frame, it has a better 'view' of the accretion shock. This
is particularly true for V834~Cen because, as we have shown in section
3, the accretion shock points ahead in orbital phase. In addition, the
Doppler maps show quite clearly that the inner part of the ballistic
stream is a dominant source of HeII line emission. This too will
irradiate the leading face of the secondary, but not the trailing face
because this side will be shielded by the secondary
itself. Furthermore, as we have already discussed, the earlier parts
of the ballistic stream do not appear to be magnetically influenced,
hence the accretion curtain, if present, probably does not form until
further down the stream and is therefore less likely to act as a
shield of the irradiating source.

\begin{figure*}
\psfig{file=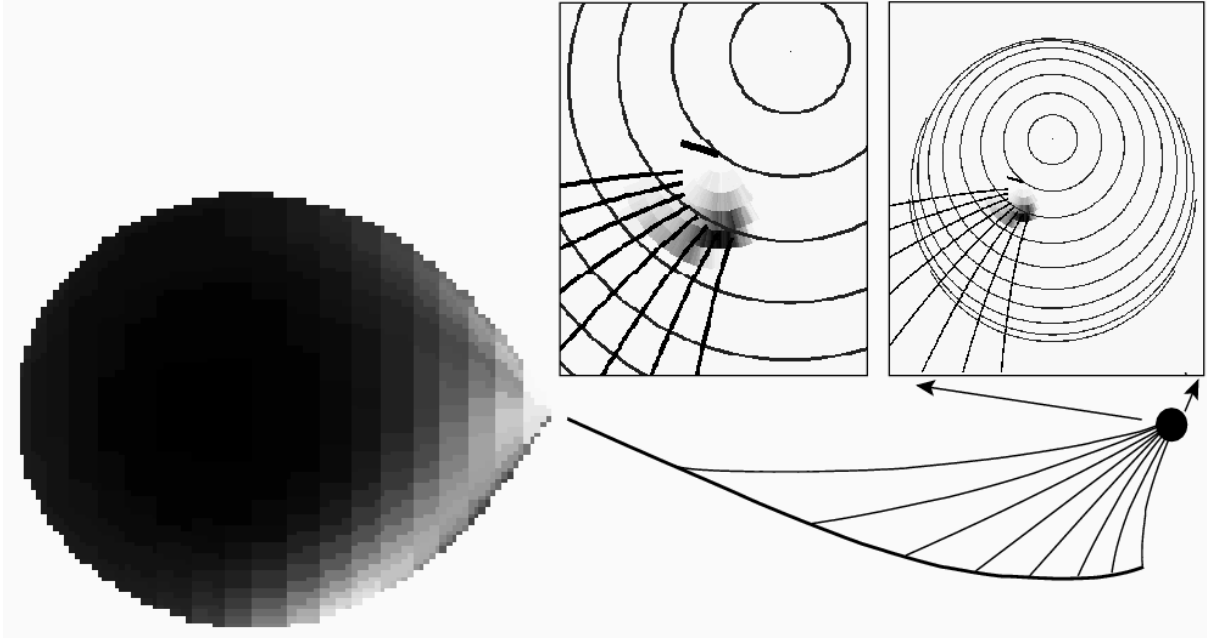,width=16cm,angle=-90}
\caption{A schematic view, from above the orbital plane, of the derived
geometry for V834~Cen}
\label{alltogethernow}
\end{figure*}

This is illustrated in Fig.~\ref{alltogethernow}, which shows a
schematic view from above the orbital plane of our derived geometry
for V834~Cen. The eight magnetic field lines correspond to the same
magnetic field lines drawn in the earlier Doppler maps. From the
insert one can see that most of the cyclotron emission comes from near
the footprints of the magnetic field lines that originated from the
latter parts of the ballistic stream. If indeed the denser parts of
the accretion curtain are located on these magnetic field lines, then
from Fig.~\ref{alltogethernow} one can see that they will not shield
the secondary from the white dwarf.

\section{Summary and Conclusions}

We have presented new photopolarimetric and spectroscopic observation
of the AM Herculis binary V834~Cen during a high state of
accretion. The observed high levels of polarization over the whole
orbit are modelled using Stokes imaging. This technique predicts a
single emission region, visible for the whole orbit, to be responsible
for the observed polarized variations. The location of the region on
the white dwarf is consistent with recent work on EUV observations by
Mauche (2002). Both analyses place the main emission region
approximately more than 40 degrees and up to 70 degrees in azimuth
from the line of centers of the two stars.

Our Doppler tomographic analysis shows several components associated
with the secondary star, the ballistic stream and parts of the
magnetically confined stream. Interestingly, the ballistic stream (at
least the earlier parts) do not show a reduced Doppler $v_{y}$
velocity as is commonly seen in other polars. We therefore reason that
the earlier part of the ballistic stream is relatively free from
magnetic influences of the white dwarf. Magnetic Doppler trajectories
are also calculated and shown to be generally coincident with the
magnetic signatures found in the Doppler maps. We also calculated the
location of the footprints of the magnetic field lines on the surface
of the white dwarf and showed that these are coincident with the
prediction for the location of the cyclotron emission region from
Stokes imaging.

Roche tomography of the narrow component seen in the HeII 4686 \AA \
emission line reveals two asymmetries in the emission distribution on
the Roche surface of the secondary. Namely, there is generally more
emission on the inner surface and also on the leading face of the
secondary. We argue that this is still consistent with irradiation of
the secondary, as the lack of a magnetic curtain at the beginning of
the ballistic stream reduces its obscuring ability. This is not only
supported by the Doppler tomogram, but also by Stokes imaging results,
which show that most of the accreting material comes along field lines
that intersect the later parts of the ballistic stream (see
Fig.~\ref{alltogethernow}).

\section{Acknowledgments}

CAW is employed on PPARC grant PPA/G/S/2000/00598.

\end{document}